\documentclass{aastex}
\begin{document}
  %
   
   %

     \title{
     Uniqueness in MHD in divergence form: right nullvectors and well-posedness}
     \author{Maurice H.~P.~M. van Putten}
     \affil{Massacusetts Institute of Technology, Room 2-378, Cambridge, MA 02139}
	      
		
\begin{abstract}
     Magnetohydrodynamics in divergence form describes a hyperbolic system 
     of covariant and constraint-free equations. It comprises a linear
     combination of an algebraic constraint and Faraday's equations.
     Here, we study the problem of well-posedness, and identify a preferred
     linear combination in this divergence formulation.
     The limit of weak magnetic fields shows the slow magnetosonic 
     and Alfv\'en waves to bifurcate from the contact discontinuity 
     (entropy waves), while the fast magnetosonic wave is a regular
     perturbation of the hydrodynamical sound speed.
     These results are further reported as a starting
     point for characteristic based shock capturing
     schemes for simulations with ultra-relativistic shocks in 
     magnetized relativistic fluids. 
\mbox{}\\
\mbox{}\\
{\em Subject headings:}MHD, divergence form, characteristics
\end{abstract}
  
  
	\section{Introduction}
   
   Highly relativistic astrophysical fluids have been observed as highly 
   energetic outflows, e.g.: jets in active galactic nuclei, including a 
   few optical radio-jets such as 3C273 (Pearson et al. 1981, Thomson et al.
   1993, Bhacall et al. 1995), 3C346 (Dey \& van Breugel 1994),
   M87 (Biretta et al. 1995) and PKS 1229-21 (Le Brun et al. 1996),
   microquasars in our galaxy (Hjellming \& Rupen 1995;
   Mirabel \& Rodriguez 1995; Levinson \& Blandford 1996),
   pulsar winds (Kennel \& Coroniti 1984), and fireballs 
   in recent models of $\gamma-$ray bursts (Rees \& Meszaros 1995).
   These flows are generally time-dependent, or have been
   produced in a strongly time-variable episode, and hence are 
   relativistically shocked fluid flows. In most cases, shocks are 
   responsible for brightest emission features at the highest energies.
   
   The evolution of strongly magnetized flows can be markedly different 
   from unmagnetized flows. This is already apparent from in small 
   amplitude wave-motion in ideal magnetohydrodynamics compared with
   hydrodynamics, and their distinct shock structures. The nonlinear 
   development of large scale morphology of strongly magnetized
   jets can result in features such as the formation of a nose cone
   (Clarke et al. 1986), which is absent in hydrodynamical evolution. 
   Of particular interest is the role of magnetic fiels
   in the large scale, three-dimensional stability 
   of jets and their knotted structures.
   
   Time-dependent simulations may provide the link between the 
   observed emission features and the internal structure such as
   magnetized field distribution, and boundary conditions at the source. 
   It is hoped that simulations ultimately provide constraints 
   on the flow parameters, perhaps also derived from stability criteria.
   Higher dimensional simulations of jets are performed
   by a number of groups in the approximation of 
   relativistic hydrodynamics (van Putten 1993b; Duncan \& Hughes 1994;
   Mart\'i et al. 1995, G\'omez et al. 1997) and relativistic 
   magnetohydrodynamics (van Putten 1994ab, 1996; 
   Nishikawa et al. 1997; Koide et al. 1996, 1998).  
   
   The earliest approach for time-dependent simulations on
   shocked relativistic magnetohydrodynamic flows with dynamically
   significant magnetic fields uses the equations of magnetohydrodynamics
   (MHD) in divergence form (van Putten 1991, 1993a). The divergence 
   technique obtains hyperbolic systems from partial differential-algebraic
   systems of equations, and applies more generally to the case of Yang-Mills 
   magnetohydrodynamics in SU(N) (van Putten 1994cd, Choquet-Bruhat 1994ab), 
   and general relativity (van Putten \& Eardley 1996). A linear smoothing 
   method has been used as a shock capturing scheme for this formulation
   (van Putten 1993a,1994ab,1995). Both one- and two-dimensional
   simulations on astrophysical jets are performed (van Putten 1993b, 1996, 
   Levinson \& van Putten 1997). This method is accurate and stable, and 
   generally performs well for relativistic shocked fluid flow 
   with up to moderately strong shock strengths (van Putten 1993), and 
   preserves divergence free magnetic fields to within machine round-off 
   error (van Putten 1995). A smoothing method, therefore, is appropriate for
   simulations on the large scale morphology of astrophysical jets.
   
   Advanced shock capturing schemes are commonly based on characteristics, 
   however, such as Roe's method (1981) and its extensions. 
   These methods are generally more stable than smoothing methods for flows 
   with ultra-relativistic shocks, such as in calculations of fire-balls 
   for $\gamma-$ray bursts (Rees \& Meszaros 1994; Wen $et$ $al.$ 1997). 
   It is therefore of interest to explore applications of these shock 
   capturing schemes to relativistic MHD. Here, we describe a first step in 
   this direction is given by studying the computational stability of normalized 
   right nullvectors ($i.e.,$ the right eigenvectors) of the characteristic 
   matrix.
   
   The divergence technique incorporates a constraint $c=0$ into a divergence 
   equation of a two-form, $\nabla^a\omega_{ab}=0$, as in Faraday's equation, 
   through the linear combination
   \begin{eqnarray}
   \nabla^a(\omega_{ab}+\lambda g_{ab}c)=0, ~~\lambda\ne0. 
   \label{EQN_RFT}
   \end{eqnarray}
   In the context of an Cauchy problem, (\ref{EQN_RFT}) conserves $c=0$ in the 
   future domain of dependence of the initial hypersurface with physical 
   Cauchy data (van Putten 1991).
   
   In this paper, we identify a preferred linear combination in 
   (\ref{EQN_RFT}), i.e.: a choice of $\lambda$ and overall sign of 
   (\ref{EQN_RFT})) in its application to the equations of ideal MHD.
   This follows from two separate analysis: a derivation of the right 
   nullvectors of the characteristic matrix and well-posedness.
   Somewhat remarkably, both analysis agree in their preferred linear 
   combinations. This suggests to consider this preferred linear 
   combination in future applications in characteristic based methods 
   to MHD in divergence form.

   The problem of linearized perturbations in relativistic MHD
   has been considered previously by Anile (1989) and that of Alfv\'en waves 
   by Kommissarov (1997). The well-posedness proof uses an extension to
   the Friedrichs-Lax symmetrization procedure from earlier work on
   Yang-Mills magnetohydrodynamics (van Putten, 1994cd).

   Section 2 describes non-uniqueness in the original version of the
   divergence technique. In Section 3, a new derivation of the
   right nullvectors is given. Section 4  briefly summarizes 
   well-posedness obtained by embedding of physical solutions in a 
   symmetric hyperbolic system of equations.

   \section{MHD in divergence form}

   Ideal MHD describes an inviscid, perfectly conductive plasma in a single 
   fluid description with velocity four-vector, $u^b$ ($u^cu_c=-1$). 
   It is given by energy-momentum conservation,
   $\nabla_aT^{ab}=0$, where $T^{ab}$ is the stress-energy tensor
   of both the fluid and the electromagnetic field, Faraday's equations,
   $\nabla_a(u^{[a}h^{b]})=0$ subject to the algebraic constraint $u^ch_c=0$, 
   and conservation, $\nabla_a (ru^a)=0$, of baryon number, $r$. For
   a polytropic equation of state with polytropic index $\gamma$, we have
   $T^{ab}=(r+\frac{\gamma}{\gamma-1}\frac{P}{r}+h^2)u^au^b+
   (P+h^2/2)g^{ab}-h^ah^b$, $P$ is the hydrostatic pressure and
   $g^{ab}$ is the metric tensor. The theory of relativistic
   magnetohydrodynamics is contained in the conservation laws
   of energy-momentum, $\nabla_aT^{ab}=0$, and baryon number, 
   $\nabla_a(ru^a)=0$, together with Faraday's equations and a constraint,
   \begin{eqnarray}
   \nabla_a(h^{[a}u^{b]})=0,~~~
   u^ch_c=0.
   \end{eqnarray}
   The divergence technique considers a constraint-free
   formulation by taking a linear combination
   \begin{eqnarray}
   \nabla_a(h^{[a}u^{b]}+
   \lambda u^ch_c)=0.
   \label{EQN_DIV}
   \end{eqnarray}
   Provided $\lambda\ne0$, (\ref{EQN_DIV}) preserves $u^ch_c=0$ during 
   dynamical evolution in response to physical initial data
   (van Putten 1991), and no constraint violating wave-motion occurs.
  
   Algebraically, the linear combination (\ref{EQN_DIV})
   establishes a rank-one update to its Jacobian,
   and hence of that of the full equations of MHD.
   Clearly, symmetry conditions of the Jacobian may enter
   a particular choice of $\lambda$. Below, we consider the choice 
   \begin{eqnarray}
   \lambda=1,
   \end{eqnarray}
   so that 
   \begin{eqnarray}
   \left\{
   \begin{array}{rl}
   \nabla_aT^{ab}=0,\\
   -\nabla_a(h^{[a}u^{b]}+g^{ab}u^ch_c)=0,\\
   \nabla_a(ru^a)=0,\\
   \nabla_a\left(\xi^a(u^2+1)\right)=0,
   \end{array}
   \right.
   \label{EQN_1}
   \end{eqnarray}
   where $\xi$ is any time-like vector field and $U=(u^b,h^b,r,P)$.
   The minus sign in front of the present linear combination is chosen also
   in regards to the structure of the Jacobian of (\ref{EQN_1}).
   This will be made explicit below.
   
   Upon expansion, (\ref{EQN_1}) obtains the system
   \begin{eqnarray}
   A^{a}\partial_aU+\cdots=0,
   \end{eqnarray}
   where the matrices $A^{aA}_B=A^{aA}_B(U)
   =\frac{\partial F^{aA}}{\partial U^B}$ are 10 by 10, and
   the dots refer coupling terms to the Christoffel symbols.
   The infinitesimal wave-structure is given by characteristic wave-fronts 
   at given $U$ (since the $A^a$ are coordinate independent). 
   The simple wave ansatz $U=U(\phi)$ obtains
   \begin{eqnarray}
   A^{a}\partial_a\phi U^\prime+\cdots=0.
   \label{EQN_2}
   \end{eqnarray}
   The wave-fronts are characteristic surfaces, whenever
   the matrix $A^{a}\partial_a\phi$ is singular. The directions
   $\nu_a=\partial_a\phi$ then are the normals to these 
   surfaces. The small amplitude perturbations in these
   simple waves are given by the right nullvectors of $A^a\nu_a$.  
   Stated differently, the small amplitude 
   perturbations are right eigenvectors, $R$, of $(A^t)^{-1}A^x\nu_x$,
   when the wave moves along the $x-$direction, following
   \begin{eqnarray}
   \left((A^t)^{-1}A^x-v\right)R=0,
   \end{eqnarray}
   where $v$ is the velocity of propagation.
   
   The divergence technique provides an embedding of the theory of ideal
   MHD in a system of ten equations. Physical initial data are properly 
   propagated by it, without exiting non-physical wave-modes.
   The physical waves (entropy waves, Alfv\'en and magnetohydrodynamic
   waves) are all contained within the light cone. Here, adding $g^{ab}u^ch_c$
   (or a multiple thereof) to Faraday's equations provides a 
   rank-one update to the characteristic matrix $A^c\nu_c$. 
   On the light cone, however, $\nu^2=0$, and this linear combination no 
   longer regularizes  of the characteristic determinant. 
   (This results from insisting on covariance in the 
   divergence formulation.) Kommissarov (1997) attempts 
   to discusses MHD in divergence form outside the context of the initial 
   value problem with physical initial data, and hence erroneously concludes
   the presence of non-physical wave-modes.
  
   \section{The characteristic matrix}
   We have
   \begin{eqnarray}
   A^{aA}(U)=
   \frac{\partial F^{aA}}{\partial U^B}\nu_a
   =
   \frac{\partial F^{aA}\nu_a}
   {\partial U^B};
   \end{eqnarray}
   with $\rho=
   r+\frac{\gamma}
   {\gamma-1}P+h^2=rf+h^2$, 
   we have
   \begin{eqnarray}
   F^{cA}\nu_c=\left\{
   \begin{array}{l}
   \rho (u^c\nu_c) u^a+(P+h^2/2)\nu^a-(h^c\nu_c)h^a,\\
   -\{(h^c\nu_c)u^a-(u^c\nu_c)h^a+\nu^au^ch_c\},\\
   r(u^c\nu_c),\\
   (\xi^c\nu_c)(u^2+1).
   \end{array}
   \right.
   \label{EQN_F}
   \end{eqnarray}
   The system of 10$\times$10 equations for $U^B=(u^b,h^b,r,P)$ can
   be reduced to 8$\times$8 in the variables $V^B=(v^s,h^b,r)$ 
   by expressing $u^b$ in terms of the spatial three-velocity 
   $u^b=\Gamma(1,v^s)$, $\Gamma=1/\sqrt{1-v^2}(1,v^s)$,
   $s=1,2,3$, noting that linearized wave-motion conserves entropy, so that
   $\mbox{d}P=\gamma\frac{P}{r}\mbox{d}r$. In $V^B$, the equation of
   energy conservation, $\nabla_aT^{at}=0$ and the last equation
   of (\ref{EQN_1}) are automatically satisfied, whence they can be
   ignored.  In what follows, $A^{a}$ shall denote the resulting
   8$\times$8 matrix, obtained from the original 10$\times$10 matrix by
   deletion of the first and last row, addition of the 
   last column (multiplied by $\frac{\gamma P}{r}$) to the one-but last
   column (associated with $r$), followed by deletion of the first
   and last columns.

   The linearized wave-structure is given by the characteristic problem
   \begin{eqnarray}
   A^{c}\nu_cz=0
   \label{EQN_4}
   \end{eqnarray}
   for the right null-vectors $z=U^\prime$. Without loss of generality, 
   (\ref{EQN_4}) can be studied in a co-moving frame, in which 
   $u^b=(1,0,0,0)$. In this event, $\Gamma=1$ and 
   $\frac{\Gamma}{\partial v^s}=0$. Furthermore, the $x$-axis of the
   local coordinate system can be aligned with the
   magnetic field, so that $h^b=(0,H,0,0)$.
   Given the two orientations $u^s$ and $h^b$,
   the wave-structure is rotationally symmetric
   bout the $x$-axis, and hence $\nu_y$ and $\nu_z$ act symmetrically as 
   $\sqrt{\nu_y^2+\nu_z^2}$; we will put $\nu_z=0$.
   For $A^c\nu_c$, we have
   \begin{eqnarray}
   \tiny{
\left [\begin {array}{cccccccc} \rho\,\nu_{{1}}&0&0&-\nu_{{1}}H&-H\nu_
{{2}}&-H\nu_{{3}}&0&{\frac {\gamma\,P\nu_{{2}}}{r}}
\\\noalign{\medskip}0&\rho\,\nu_{{1}}&0&0&H\nu_{{3}}&-H\nu_{{2}}&0&{
\frac {\gamma\,P\nu_{{3}}}{r}}\\\noalign{\medskip}0&0&\rho\,\nu_{{1}}&0
&0&0&-H\nu_{{2}}&0\\\noalign{\medskip}\nu_{{1}}H&0&0&-\nu_{{1}}&-\nu_{
{2}}&-\nu_{{3}}&0&0\\\noalign{\medskip}-H\nu_{{2}}&H\nu_{{3}}&0&\nu_{{
2}}&\nu_{{1}}&0&0&0\\\noalign{\medskip}-H\nu_{{3}}&-H\nu_{{2}}&0&\nu_{
{3}}&0&\nu_{{1}}&0&0\\\noalign{\medskip}0&0&-H\nu_{{2}}&0&0&0&\nu_{{1}
}&0\\\noalign{\medskip}r\nu_{{2}}&r\nu_{{3}}&0&0&0&0&0&\nu_{{1}}
\end {array}\right ]
}.
   \label{EQN_FM}
   \end{eqnarray}
   Note that the lower diagonal block is $\nu_1$ times the
   4$\times$4 indentity matrix. {\em This results from the
   sign choice in the given combination of Faraday's 
   equations and the constraint in (\ref{EQN_1}) and (\ref{EQN_F}).} 
   Furthermore, notice that the third and seventh rows and
   columns act independently to give rise to the Alfv\'en waves.
   The remaining waves are described by the reduced problem
   \begin{eqnarray}
   (A^{c}\nu_c)^\prime z^\prime=0,
   \label{EQN_4B} 
   \end{eqnarray}
   where $(A^c\nu_c)^\prime$ is obtained from $A^c\nu_c$ by deleting
   the third and seventh rows and columns, thereby obtaining a problem 
   in the 6-dimensional variable $z^\prime$. Introducing
   \begin{eqnarray}
   z^\prime=\left(
   \begin{array}{c}x\\y\end{array}
   \right),
   \end{eqnarray}
   (\ref{EQN_4}) takes the form of a coupled system of 3$\times$3 equations
   \begin{eqnarray}
   \nu_1Zx+Xy=0,~~Yx+\nu_1y=0,
   \label{EQN_C}
   \end{eqnarray}
   in which
   \begin{eqnarray}
   \begin{array}{llc}
   Z=
\left [\begin {array}{ccc} \rho&0&-H\\\noalign{\medskip}0&\rho&0
\\\noalign{\medskip}H&0&-1\end {array}\right ]
   ,\\
   \mbox{}\\
   X=
\left [\begin {array}{ccc} -H\nu_{{2}}&-H\nu_{{3}}&{\frac {\gamma\,P
\nu_{{2}}}{r}}\\\noalign{\medskip}H\nu_{{3}}&-H\nu_{{2}}&{\frac {
\gamma\,P\nu_{{3}}}{r}}\\\noalign{\medskip}-\nu_{{2}}&-\nu_{{3}}&0
\end {array}\right ]
   ,\\
   \mbox{}\\
   Y=
\left [\begin {array}{ccc} -H\nu_{{2}}&H\nu_{{3}}&\nu_{{2}}
\\\noalign{\medskip}-H\nu_{{3}}&-H\nu_{{2}}&\nu_{{3}}
\\\noalign{\medskip}r\nu_{{2}}&r\nu_{{3}}&0\end {array}\right ]
   .
   \end{array}
   \end{eqnarray}
   This obtains a $single$ 3$\times$3 eigenvalue problem in $x$,
   given by
   \begin{eqnarray}
   XYx=\nu_1^2Zx~~\Leftrightarrow
   ~~Z^{-1}XYx=\nu_1^2x.
   \end{eqnarray}
   Here, $Z^{-1}XY-\nu_1^2$ is given by the matrix
   \begin{eqnarray}
   \begin{array}{ccc}
   \tiny{
\left [\begin {array}{ccc} W_{{1,1}}&W_{{1,2}}&0\\\noalign{\medskip}W_{{2,1}}&W_{{2,2}}
&0\\\noalign{\medskip}{\frac {H\left (\gamma\,P{\nu_{{2}}}^{2}-{\it rf}\,{\nu_{{2}}}^{2
}-{\it rf}\,{\nu_{{3}}}^{2}\right )}{{\it rf}}}&{\frac {H\gamma\,P\nu_{{2}}\nu_{{3}}}{{
\it rf}}}&{\nu_{{2}}}^{2}+{\nu_{{3}}}^{2}-{\nu_{{1}}}^{2}\end {array}\right ]
   }
   \end{array}
   \label{EQN_ZIXY}
   \end{eqnarray}
   where the upper diagonal 2$\times$2 matrix $W$ is given by
   \begin{eqnarray}
   W=%
{\it Wij}
   . 
   \label{EQN_W}
   \end{eqnarray}
   {\em The two zeros in the third column of (\ref{EQN_ZIXY}) result
   from $\lambda=1$.} Upon substitution $\nu_3^2=\nu^2+\nu_1^2-\nu_2^2$, 
   the determinant assumes the covariant expression
   \begin{eqnarray}
   \begin{array}{rl}
   \rho
   ~\mbox{det}W=&
   (rf-\gamma P)
   (u^c\nu_c)^4\\
   &-(h^2+\gamma P)
   \nu^2(u^c\nu_c)^2
   +\frac{\gamma P}{rf}(h^c\nu_c)^2\nu^2.
   \end{array}
   \label{EQN_DET}
   \end{eqnarray}
   {\em Alfv\'en waves.} The eigenvalues for the Alfv\'en waves are 
   given by
   \begin{eqnarray}
   \nu_1
   =\pm \frac{|h^c\nu_c|}
   {\sqrt{\rho}}
   \end{eqnarray}
   with null-vector
   \begin{eqnarray}
   z=(0,0,H\nu_2,0,0,0,\rho\nu_1,0)^T, 
   \end{eqnarray}
   associated with Alfv\'en waves; covariantly,
   \begin{eqnarray}
   U^A=(v^a,\pm\sqrt{\rho}v^a,0,0)^T,
   \label{EQN_AL}
   \end{eqnarray}
   where $v_a$ may be taken to be
   \begin{eqnarray}
   H(0,0,\nu_4,-\nu_3)=
	  \epsilon_{abcd}
	  u^bh^c\nu^d\equiv v_a.
   \label{EQN_VA}
   \end{eqnarray}
   Thus, the Alfv\'en wave is transversal in which 
   $h^2$ is conserved ($\delta h^b$ is orthogonal to $h^b$).
   
   {\em Magnetohydrodynamic waves.} The eigenvalues for the
   magnetohydrodynamic waves are given by the roots
   of the characteristic determinant
   (\ref{EQN_DET}). Writing 
   \begin{eqnarray}
   n^b=\nu^b+(u^c\nu_c)u^c,
   \label{EQN_NN}
   \end{eqnarray}
   we have
   $\nu^2=-t^2+n^2,~~t=u^c\nu_c,~~n^2=n^cn_c.$ 
   Let $\alpha=\frac{rf}{\gamma P}$
   and $\beta=\frac{h^2}{\gamma P}$.
   Then
   \begin{eqnarray}
   \frac{(h^c\nu_c)^2}{rfn^2}=
   \frac{\beta}{\alpha}\frac{(h^cn_c)^2}{h^2n^2}
   \equiv\frac{\beta}{\alpha}\cos^2\phi.
   \end{eqnarray}
   Consequently,  
   (\ref{EQN_DET}) becomes
   \begin{eqnarray}
   \begin{array}{rl}
   (\alpha-1)v^4-&(1+\beta)v^2(1-v^2)
   \\
   &+\beta\alpha^{-1}\cos^2\phi(1-v^2)=0, 
   \end{array}
   \label{EQN_DETB}
   \end{eqnarray}
   where $v^2=\frac{t^2}{n^2}$. (\ref{EQN_DETB})
   has real solutions $v$ for any given $n^b$, whenever
   \begin{eqnarray}
   (\alpha+\beta)v^4
   -(1+\beta+\beta\alpha^{-1})v^2
   +\beta \alpha^{-1}=0
   \label{EQN_NU}
   \end{eqnarray}
   has real solutions $v$.
   But (\ref{EQN_NU}) has discriminant
   \begin{eqnarray}
   D=(\alpha+\beta-\alpha\beta)^2\ge0.
   \end{eqnarray}
   Weak magnetic fields are described by small $\beta$ expansions as
   follows.\\
   \mbox{}\\
   {\sc Proposition 3.1.} {\em Fast magnetosonic waves
   are a regular perturbation of sound waves in pure
   hydrodynamics, while the Alfv\'en and slow magnetosonic waves
   bifurcate from entropy waves (contact discontinuities),
   whose propagation velocities satisfy}
   \begin{eqnarray}
   \begin{array}{rl}
       v^2_f/v_h^2&\sim
       1+\beta 
       \frac{\alpha-1}{\alpha}
       \sin^2\phi+O(\beta^2),\\
       \mbox{}\\
       v^2_A/v_h^2&\sim \beta\cos^2\phi[1-\beta\alpha^{-1}+O(\beta^2)],\\
       \mbox{}\\
       v^2_s/v_h^2&\sim \beta\cos^2\phi[1-\beta
		        (1-\frac{\alpha-1}{\alpha} 
			 \cos^2\phi)+O(\beta^2)],
   \end{array}
   \end{eqnarray}
   {\em where $v^2_h=\alpha^{-1}$ is the square of 
   the hydrodynamical velocity, and which obey the inequalities}
   \begin{eqnarray}
   v^2_s \le v_A^2\le v_f^2.
   \label{EQN_VVV}
   \end{eqnarray}
   Inequalities (\ref{EQN_VVV}) remain valid for general $\beta$
   ($e.g.$ Bazer \& Ericson 1959;  Lichnerowicz 1967;  Anile 1989).
   \section{Right nullvectors} Inspection of  (\ref{EQN_W}), together
   with (\ref{EQN_C}), shows the  null-vector
   \begin{eqnarray}
   z=
   \left(\begin{array}{c}
   \nu_1\nu_2\nu_3^2\\
   -\nu_1\nu_3(\nu_2^2-\alpha\nu_1^2)\\
   0\\
   H\nu_1\nu_2\nu_3^2\\
   H\nu_3^2(\nu_2^2-\alpha\nu_1^2)\\
   -H\nu_2\nu_3(\nu_2^2-\alpha\nu_1^2)\\
   0\\
   -\alpha r\nu_3^2\nu_1^2
   \end{array}
   \right).
   \label{EQN_Z}
   \end{eqnarray}
   Of course, (\ref{EQN_Z}) can be stated covariantly
   by noting that $H^2=h^2$, $H\nu_2=h^c\nu_c$,  $\nu_1=u^c\nu_c$,
   \begin{eqnarray}
   H^2(\nu_2^2-\alpha\nu^2_1)
   =(h^c\nu_c)^2-\alpha
   h^2(u^c\nu_c)^2\equiv h^2k_1,
   \end{eqnarray}
   and introducing
   \begin{eqnarray}
   H(0,\nu_4^2+\nu_3^2,
   -\nu_2\nu_3,-\nu_2\nu_4)^T&=&
   \epsilon_{abcd}u^b\nu^c v^d\equiv w_a.
   \label{EQN_WA}
   \end{eqnarray} 
   Since $-\alpha r\nu_3^2\nu_1^2$  is a scalar,  $\nu^3$ is to be treated as
   \begin{eqnarray}
   H^2(\nu_3^2+\nu_4^2) =h^2n^2-(h^c\nu_c)^2 \equiv h^2k_2,
   \end{eqnarray}
   were $n_a=\nu_a+(u^c\nu_c)u_a$.  Note that
   \begin{eqnarray}
   k_1=n^2(\cos^2\phi-\alpha v^2), ~~~k_2=n^2\sin^2\phi,
   \end{eqnarray}
   where $v=v_s, v_f$.  Clearly, $z$ is formed from
   \begin{eqnarray}
   \begin{array}{rl}
   \delta u^b&=-t( k_1n^b-(k_2+k_1) (\hat{h}^c n_c)\hat{h}^b)\\
   \delta h^b&=k_1w^b+k_2 t(h^c n_c)u^b,\\
   \delta r &=  -\alpha rk_2t^2,\\
   \delta P &=-rfk_2t^2,
   \end{array}
   \label{EQN_REIG0}
   \end{eqnarray}
   where $\hat{h}^b=h^b/|h|$, and
   \begin{eqnarray}
   v_a=\epsilon_{abcd}u^bh^c\nu^d,~~~ w_a=\epsilon_{abcd}u^b\nu^c v^d.
   \end{eqnarray}
   We thus have  the following.
   \mbox{}\\
   \mbox{}\\
   {\sc Proposition 3.2.} {\em Given a unit vector
   $n^b$ orthogonal to $u^b$, and a root
   $\nu^b=n^b+vu^b$, $v=u^c\nu_c$ of (\ref{EQN_NU}),
   the right nullvectors for  the hydrodynamical 
   waves of (\ref{EQN_4}),  $U^A=(\delta u^b, \delta h^b,\delta r,
   \delta P)$,  are} 
   \begin{eqnarray}
   \begin{array}{rl}
   \delta u^b&=  ~v\left[\sin^2\phi~n^b -(1-\alpha v^2)
   (n^b-\cos\phi~\hat{h}^b)\right],\\
   \mbox{}\\
   \delta h^b&=
       |h|\left[ (\cos^2\phi-\alpha v^2)\tilde{w}^b
      +v\sin^2\phi\cos\phi  ~u^b\right],\\
   \mbox{}\\
   \delta r &=  -v^2\alpha r\sin^2\phi,\\
   \mbox{}\\
   \delta P &= -v^2 rf\sin^2\phi.
   \end{array}
   \label{EQN_REIG}
   \end{eqnarray}
   {\em where $\tilde{w}^b=w^b/|h|$.}\\
   \mbox{}\\
   Anile (1989) gives  a different form of these
   right nullvectors.   By Proposition 3.1, 
   our weak magnetic field limits  show that
   \begin{eqnarray}
   \cos^2\phi-\alpha v_f^2<0
   \end{eqnarray}
   for fast magnetosonic waves,  while
   \begin{eqnarray}
   \cos^2\phi-\alpha v_s^2>0
   \end{eqnarray}
   for slow magnetosonic waves. Inspection of (\ref{EQN_WA})
   shows that therefore the tangential component of the magnetic
   field is strengthened in fast magnetosonic waves,
   while it is weakened in slow magnetosonic
   waves. This distinguishing aspect of fast and slow
   magnetosonic waves was first noted  by Bazer \& Ericson (1959) in their
   analysis of shocks in nonrelativistic MHD.
   
   The limit of small $\beta$ is of particular interest to computation.
   For example, in various settings  a magnetized fluid streams into
   a nearly unmagnetized environment. A characteristics based scheme
   must therefore reliable treat  a large dynamic range in
   $\beta$. Clearly, a full set of nullvectors (including those of
   contact discontinuities) obtains for nonzero $\beta$. However,
   the behavior of these nullvectors is somewhat nontrivial as $\beta$
   becomes small. In what follows, we consider the small $\beta$
   limit, in the sense of small $|h|/\sqrt{\gamma P}$,
   while keeping the direction $\hat{h}^b$ constant. In this limit,
   \begin{eqnarray}
   \begin{array}{rl}
   1-\alpha v^2&\sim -\beta  \frac{\alpha-1}{\alpha}
		 \sin^2\phi+O(\beta^2),\\
   1-\alpha v^2&\sim 1+O(\beta)
   \end{array} 
   \end{eqnarray} 
   for the fast and slow magnetosonic speeds, respectively.\\
   \mbox{}\\
   {\sc Corollary 4.1.} 
   {\em In the limit of
   low magnetic field strength, the fast magnetosonic waves
   are described by the right nullvectors}
   \begin{eqnarray}
   \begin{array}{ll}
   \delta u^b&= v_fn^b+\beta \frac{\alpha-1}{\alpha}
       (n^b-\cos\phi\hat{h}^b)v_f+O(\beta^2),\\
   \delta h^b&=|h|(-\tilde{w}^b+v_f\cos\phi~u^b)
               +\beta\frac{\alpha-1}{\alpha}w^b +O(\beta^{2}),\\
   \delta r &= -v_f^2\alpha r,\\
   \delta P &= -v_f^2 rf,
   \end{array}
   \label{EQN_REIGC}
   \end{eqnarray}
   {\em and the slow magnetosonic waves by}
   \begin{eqnarray}
   \begin{array}{ll}
   \delta u^b&= \cos\phi (\hat{h}^b- \cos\phi n^b)+O(\beta),\\
   \delta h^b&=\sqrt{\gamma P} (\cos\phi~\tilde{w}^b
   +v_s\sin^2\phi~u^b) +O(\beta),\\
   \delta r &=  -v_s\alpha r\sin^2\phi,\\
   \delta P &= -v_s\alpha rf\sin^2\phi.
   \end{array}
   \label{EQN_REIGB}
   \end{eqnarray}
   The small $\beta$ limit of the nullvectors can now be normalized.
   
   \subsection{Bifurcations  from  entropy waves}
   
   The behavior of the nullvectors in the limit of weak magnetic fields 
   can be derived from (\ref{EQN_AL}) and Corollary 4.1.
   To this end, note that
   \begin{eqnarray}
   v^a=|h|\tilde{v}^a=\sin\phi~|h|\hat{v}^a,
   \end{eqnarray}
   where $\hat{v}^c\hat{v}_c=1$, and $\phi$ denotes the angle
   between $n^c$ and $h^c$,
   \begin{eqnarray}
   n^b=\cos\phi~\hat{h}^b+\sin\phi~y^b,
   \end{eqnarray}
   $y^cu_c=h^cy_c=0, y^cy_c=1$ ($n^b$ is normalized to be
   unit, as in the assumptions of Proposition 3.2). 
   It follows that the Alfv\'en nullvectors may be
   normalized to
   \begin{eqnarray}
   \delta \hat{U}^A=(\hat{v}^a,\pm\sqrt{\rho}\hat{v}^a,0,0).
   \label{EQN_AA}
   \end{eqnarray}
   In the limit of  vanishingly small $\beta$,
   the pair of slow magnetosonic waves 
   collapse to the single normalized nullvector
   \begin{eqnarray}
   \delta\hat{U}^A=(y^b,\sqrt{\gamma P}y^b,0,0). 
   \label{EQN_BB}
   \end{eqnarray}
   Note that $y^c\hat{v}_c=0$, so that
   (\ref{EQN_AA}) and (\ref{EQN_BB}) are
   independent. Division by $\sin\phi$ thus provides a
   normalization of the original expressions
   (\ref{EQN_AL}) and (\ref{EQN_REIGB}).
   
   The nullvector associated with entropy waves
   ($u^c\nu_c=0$) is
   \begin{eqnarray}
   \delta U^A=
   (0,0,\delta r,0)
   \end{eqnarray}
   if $h^c\nu_c\ne0$, and
   \begin{eqnarray}
   (0,\delta h^c,\delta r,\delta P),
   ~~(\delta u^c,0,0,0),
   \end{eqnarray}
   if $h^c\nu_c=0$, subject to
   \begin{eqnarray}
   \delta P+h_c\delta h^c=0,~~\nu_c\delta h^c=0,~~\nu_c\delta u^c=0.
   \end{eqnarray}
   The second case refers to transverse MHD for 
   which continuity must hold of total pressure, zero orthogonal magnetic
   field and transverse velocity. Note that transverse MHD
   has two nullvectors, and corresponds to the case of
   pure hydrodynamics. With the exception of transverse MHD, therefore,
   the contact discontinuity provides one nullvector.

   Transverse MHD or pure hydrodynamics allows for shear along 
   contact discontinuities, which is responsible for
   the two independent nullvectors. Whenever 
   magnetic field lines cross a contact discontinuity, however,
   the persistent coupling to the magnetic field lines in ideal MHD 
   prohibits shear. In ideal MHD, the response to the original
   two-dimensional degree of freedom in shear is two new wave-modes. 
   These wave modes are the Alfv\'en wave and the slow magnetosonic wave.
   These two wave-modes are indeed different, as (\ref{EQN_AA}) and
   (\ref{EQN_BB}) show. The Alfv\'en and slow
   magnetosonic wave may be regarded as one pair,
   bifurcating from the contact discontinuity. This has been illustrated
   in Fig. 6 of van Putten (1993a). Indeed, {\em the limit
   of vanishing $\beta$ recovers the two shear modes from
   the independent Alfv\'en and slow magnetosonic waves.}
   Of course, the Alfv\'en wave is purely rotational, while
   the slow magnetosonic wave is slightly helical, including
   a longitudinal variation of $\pm v_s\sin^2\phi=
   \pm \beta\sin^2\phi\cos\phi$. The fast magnetosonic wave
   remains a regular perturbation of the ordinary sound wave.
   
   The weak magnetic field limit thus obtains two nullvectors
   from the fast magnetosonic waves, two from the Alfv\'en waves,
   one from the slow magnetosonic waves and generally one from 
   the contact discontinuity, a total of six. This leaves an
   apparent degeneracy of one.

   The degeneracy stems from the neighboring to
   order $v_s$ of the two nullvectors of the slow
   magnetosonic waves. This would suggest
   ill-posedness to this order in projections.
   However, characteristic based methods consider
   the product of the projections on the
   nullvectors $and$ the associated eigenvectors.
   In the present case, therefore, the order of the degeneracy
   is precisely cancelled by multiplication
   with the eigenvalue $v_s$, which is computationally stable.
   The limit of arbitrarily small $\beta$ {\em in the application
   of characteristic based methods} is computationally well-posed.

   \section{Well-posedness}
   The theory of ideal relativistic MHD was first shown to be well-posed by 
   Friedrichs (1974), using the Friedrichs-Lax symmetrization procedure
   (1971). The problem of constraints was circumvented by a reduction of 
   variables. The symmetrization procedure of Friedrichs and Friedrichs 
   and Lax (1971) applies to hyperbolic systems of equations of the form
   \begin{eqnarray}
   \nabla_aF^{aB}=f^B 
   \label{EQN_FL}
   \end{eqnarray}
   which satisfy a certain convexity condition. The presence of conserved 
   constraints, however, can be treated also by an extension of the 
   Friedrichs-Lax symmetrization procedure, with no need for an 
   additional reduction of variables, developed in earlier work on Yang-Mills
   magnetohydrodynamics in SU(N) (van Putten 1994cd). Once in symmetric 
   hyperbolic form, well-posedness results from standard 
   energy arguments (e.g. Fisher \& Marsden 1972).
   The main arguments of symmetrization
   in the presence of constraints are briefly recalled here, to 
   highlight the same linear combination of (\ref{EQN_1}),
   now from the point of view of well-posedness.
   
   \subsection{Symmetrization with constraints}
   Variations $\delta V^A$ of $(u^b,h^b,r,P)$ can be
   unconstraint (with respect to all ten degrees of freedom),
   and constraint, i.e., those obeying the constraints. For example,
   $\delta c\ne0$ results from a total variation, while $\delta c=0$ is 
   a constraint variation. Symmetrization in the presence of
   constraints follows if there exists a vector field $W_A$ which produces
   a total derivative in the modified main dependency relation
   \begin{eqnarray}
   \mbox{YI}:~~W_A\delta F^{aA}\equiv\delta z^a,
   \end{eqnarray}
   and which obtains constraint positive definiteness in
   \begin{eqnarray}
   \mbox{YII}:~~\delta W_A\delta F^{aA}\xi_a > 0 
   \end{eqnarray}
   for some time-like vector $\xi^a$. Of course, the source terms
   $f^B$ must satisfy the consistency condition
   \begin{eqnarray}
   W_Af^A=0
   \end{eqnarray}
   whenever the constraints are satisfied. Allowing a possible
   nonzero total derivative in YI defines an extension
   (van Putten 1994cd) to the Friedrichs-Lax (1971)
   symmetrization procedure.
   
   Differentiation by $V^C$ of the unconstraint identity  YI obtains
   \begin{eqnarray}
   \frac{\partial W_A}{\partial V^C}
   \frac{\partial F^{aA}}{\partial V^D}
   \nabla_a V^D
   + \frac{W_A\partial^2 F^{aA}}
   {\partial V^C\partial V^D}
   \nabla_aV^D=
   \frac{\partial^2 z}{\partial V^C\partial V^D} 
   \nabla_aV^D.
   \end{eqnarray}
   This establishes symmetry of the matrices 
   \begin{eqnarray}
   A^{a}_{CD}=
   \frac{\partial W_A}{\partial V^C}
   \frac{\partial F^{aA}}{\partial V^D}
   \end{eqnarray}
   Also,
   \begin{eqnarray}
   \delta V^CA^{a}_{CD}\xi_a\delta V^D&=
   (\delta V^C\frac{\partial W_A}{\partial V^C})
   (\frac{\partial F^{aA}\xi_a}{\partial V^D}\delta V^D)
   = \delta W_A \delta F^{aA}\xi_a >0
   \end{eqnarray}
   for all constraint variations 
   $\delta V^A$. Of course, given $V^A$, the constraint
   variations $\delta V^A$ define a linear
   subspace ${\cal V}$ of dimension $N-m$, where $m$
   is the number of constraints $c=0$, each giving rise to
   \begin{eqnarray}
   0=\delta c=\frac{\partial c}{\partial V^A}\delta V^A.
   \end{eqnarray}
   We have the following construction (van Putten, 1994cd).
   \vskip.1in
   {\sc Lemma 5.1} {\em Given a real-symmetric
   $A\epsilon{\cal L}(\mbox{\sc{R}}^n, \mbox{\sc{R}}^n)$ which is positive 
   definite on a linear subspace ${\cal V}\subset \mbox{\sc{R}}^n$, there 
   exists a real-symmetric, positive definite $A^*\epsilon {\cal L}
   (\mbox{\sc{R}}^n,\mbox{\sc{R}}^n)$ such that}
   \begin{eqnarray}
   A^*y=Ay~~(y\epsilon{\cal V}). 
   \end{eqnarray}
   This may be seen as follows. Consider $A^*=A+\mu x^Tx$, where
   $x$ is a unit element from $V^\perp$. Then $A^*$ is symmetric positive definite
   on $V^\prime=\{z=y+\lambda x|y \epsilon V,\lambda \epsilon R\}:$ 
   $z^TA^Tz \ge c^\prime
   ||z||^2=c^\prime(||y||^2+\lambda^2 ||x||^2)$ with $c^\prime>0$ upon choosing 
   $\mu>M$, where
   $M=||A||$ denotes the norm of $A$. This construction may be repeated until
   $V^\perp$ is exhausted, leaving $A^*$ symmetric positive definite on 
   $\mbox{\sc{R}}^n$
   as an embedding of $A$ on $V$.

   The real-symmetric matrix $A^{a}_{CD}\xi_a$  is positive
   definite on the subspace of constraint variations ${\cal V}$;
   let $(A^{a}_{CD}\xi_a)^*$ be the positive definite, symmetric
   matrix obtained from the Lemma. It follows that solutions to
   (\ref{EQN_FL}) (and its constraints) satisfy the {\em symmetric 
   positive definite} system of equations
   \begin{eqnarray}
   -(A^{aAB})^*\xi_a(\xi^c\nabla_c)V_A+ A^{a AB}(\nabla_\Sigma)_a V_A=f^B,
   \label{EQN_SE}
   \end{eqnarray}
   where 
   \begin{eqnarray}
   \nabla_a=-\xi_a(\xi^c\nabla_c)+(\nabla_\Sigma)_a. 
   \end{eqnarray}
   It remains to show that ideal MHD satisfies properties YI
   and YII.
   
   \subsection{Symmetrization of hydrodynamics}

   Relativistic hydrodynamics has been shown to be symmetrizable
   by Friedrichs (1974), Ruggeri \& Strumia (1981), and Anile (1989). 
   This uses the equations in the form
   \begin{eqnarray}
   \nabla_a F_f^{aA}=\left\{
   \begin{array}{rl}
   \nabla_a(rfu^au^b+Pg^{ab})&=0,\\ 
   \nabla_a(ru^a)&=0,\\
   \nabla_a(rSu^a)&=0
   \end{array}
   \right.
   \label{EQN_HYD2}
   \end{eqnarray}
   away from entropy generating shocks. Then $W_A^f=(u_a,f-TS,T)$ and
   $V_C^f=(v_\alpha,T,f)$ with a reduction of variables on
   the velocity four-vector by $u^b=\Gamma(1,v^\alpha)$,
   where $\Gamma$ is the Lorentz factor. 
   With $F_f^{aA}$ denoting the fluid dynamical equations
   $\nabla_aT^{ab}_f=0$, $T_f^{ab}=rfu^au^b +Pg^{ab}$ 
   with $f$ the specific enthalpy, and $\nabla_a(ru^a)=0$,
   it has been shown that (Ruggeri \& Strumia 1981; Anile 1989) 
   \begin{eqnarray}
   W_A^f\delta F^{aA}_f\equiv0,~~
   Q_f=\delta W_A\delta F^{aA}_f\xi_a>0
   \end{eqnarray}
   provided that  the free enthalpy $G(T,P)=f-TS-1$ 
   is concave, and the sound velocity is less than the speed of light.
   Under this conditions, the hydrodynamical equations alone,
   therefore, satisfy YI and YII, and
   in fact the original Friedrichs-Lax conditions
   CI and CII of Friedrichs \& Lax (1971), so
   that they satisfy a symmetric hyperbolic
   system of equations.
   \subsection{Symmetrization of ideal MHD}
   In what follows, we set
   \begin{eqnarray}
   \begin{array}{rll}
   \omega_{ab}&=&h^au^b-u^ah^b+g^{ab}u^ch_c,\\
    T_m^{ab}&=&h^2u^au^b+\frac{1}{2}h^2g^{ab} -h^ah^b.
   \end{array}
   \end{eqnarray}
   We then have the expansions
   \begin{eqnarray}
   \begin{array}{rll}
   u_b\delta T^{ab}_m&=&
   u_b(h^2u^a\delta u^b+h^2u^b\delta u^a
   +2u^au^bh_c\delta h^c\\
   &&+g^{ab}h_c\delta h^c-h^a\delta h^b
   -h^b\delta h^a)\\
   &=&-h^2\delta u^a-u^a(h_c\delta h^c)
   -h^a(u_c\delta h^c)\\
   &&-c\delta h^a,\\
   h_b\delta\omega^{ab}&=&h_b(h^a\delta u^b+u^b\delta h^a-h^b\delta
   u^a-u^a\delta h^b\\
   &&+g^{ab}\delta c)\\
   &=&h^a(h_c\delta u^c)+c\delta h^a-h^2\delta u^a-u^a(h_c\delta
   h^c)\\
   &&+h^a\delta c.
   \end{array}
   \end{eqnarray}
   We hereby arrive at the identity
   \begin{eqnarray}
   u_b\delta T^{ab}_m-
   h_b\delta\omega^{ab}\equiv\delta z^a,
   \label{EQN_TD}
   \end{eqnarray}
   where $z^a=-2h^ac$. {\em The total derivative
   in (\ref{EQN_TD}) follows by the unique 
   linear combination $\omega^{ab}=h^au^b-h^bu^a+g^{ab}c$, 
   as in
   (\ref{EQN_1}).}
   With $W_A=(u_a,h_a,f-TS,S)$ and $F^{aA}$ given by
   (\ref{EQN_1}) [rewritten according to (\ref{EQN_HYD2})], it
   follows that 
   \begin{eqnarray}
   W_A\delta(F^{aA}_f+F^{aA}_m)\equiv\delta z^a.
   \end{eqnarray}
   A similar calculation
   (van Putten 1994cd)
   shows that quadratic of constraint 
   variations $Q_m$ given by
   \begin{eqnarray}
   \begin{array}{rll}
   \delta u_b\delta T^{ab}_m\xi_a&-&\delta h_b\delta\omega^{ab}\xi_a
   =(u^c\xi_c)[h^2(\delta u)^2+(\delta h)^2]\\
    &+&2[(\xi_c\delta u^c)(h_c\delta h^c)-(h^c\xi_c)(\delta u_c\delta h^c)]
   \end{array}
   \label{EQN_QM}
   \end{eqnarray}
   is positive definite (for $\delta h^a\ne0$).
   Therefore, the sum
   \begin{eqnarray}
   Q=\delta W_A\delta F^{aA}\xi_a=Q_f+Q_m
   \end{eqnarray}
   is constraint positive definite, whenever $Q_f$ is
   such (with respect to the fluid dynamical variables).
   It follows that both YI and YII are satisfied
   (with $W_A=(u_a,h_a,f-TS,S)$ and $V_A=(v_\alpha,h_a,T,f)$),
   and hence physical solutions to (\ref{EQN_1}) satisfy
   the symmetric hyperbolic system (\ref{EQN_SE}) with $f^B=0$.
   
\mbox{}\\
\centerline{\bf Acknowledgements}
\mbox{}\\
The author greatfully acknowledges A. Edelman for stimulating
discussions, and A. Levinson, T. Jones and
D. Ryu for their interest in the present work.
This work received partial support from the MIT Sloan/Cabot
Fund, and NASA Grant NAG5-7012.
\mbox{}\\
\mbox{}\\
\baselineskip10pt
\centerline{\bf References}
\mbox{}\\
\small{
Anile A.M., 1989, {\em Relativistic fluids and magneto-fluids},
        \mbox{  }\mbox{  }(Cambridge Univ. Press, Cambridge)\\
Bahcall J.N., Kirhakos S., Schneider D.P., Davis R.J., Muxlow T.W.B., Garrington
	S.T.,\\
  Conway R.G. \& Unwin S.C., 1995, ApJ Lett., 452(2), L91\\
Bazer J. \& Ericson W.B. , 1959, ApJ, 129, 758\\ 
Birretta J.A., Zhou F. \& Owen F.N., 1995, ApJ, 447, 582\\
Choquet-Bruhat, Y., 1994a, in Proc.~VIIth Conf. Waves and Stability in \\
	\mbox{  }\mbox{  }Continuous media, Bologna, 1993,
	ed. T. Ruggeri (World Scientific, Italy) \\
Choquet-Bruhat, Y., 1994b, C.R. Acad. Sci. Paris, S\'er. I Math., 318, 775\\
Clarke D.A., Norman M.L. \& Burns J.O., 1986, ApJ Lett., 311, L63\\ 
Dey A. \& van Breugel W.J.M., 1994, AJ, 107(6), 1977\\
Duncan G.C. \& Hughes P.A. 1994, ApJ, 436, L119\\
Fischer A.E. \& Marsden J.E., 1972, Commun. Math. Phys., 28, 1\\
Friedrichs K.O., 1974, Commun. Pure Appl. Math., 28, 749\\
Friedrichs K.O. \& Lax P.D., 1971, Proc. Natl. Acad. Sc. USA, 68, 1686\\
G\'omez, J.L., Mueller E., Font J.A., Ibanez J.M.A., Marquina
   \mbox{  }\mbox{  }A., 1997, ApJ. 479, 151\\
Hjellming R.M. \& Rupen M.P, 1995, Nature, 375(8), 464\\
Kennel C.F. \& Coroniti F.V., 1984, ApJ, 283, 694\\
Koide S., Nishikawa K., Mutel R.L., 1996, ApJ, 463, L71\\
Koide S., Shibata K., Kudoh T., 1998, ApJ, 495, L63\\
Kommissarov S.S., 1997, Phys. Lett. A., 232, 435\\
Le Brun V., Bergeron J., Boiss\'e P. \& Deharveng J.M., 1997, {\em A\&A}, 321, 733.\\
Levinson A.\& Blandford R.D., 1996, ApJ Lett., 456(1), L29\\	
Levinson A.\& van Putten M.H.P.M., 1997, {\em Ap.J.} 488, 69.\\
Lichnerowicz A., 1967, {\em Relativistic hydrodynamics and
magnetohydrodynamics} (W.A. Benjamin Inc., New York). \\
Lind K.R., Payne D.G., Meier D.L. \& Blandford R.D.,
   1989, ApJ, 344, 89\\
Mart\'i J.M$^a$, M\"uller E., Font J.A. \& Ib\'a\~nez J.M$^a$,
   1995, ApJ, \mbox{  }\mbox{  }448, L105\\
Mirabel I.F. \& Rodriguez L.F., 1996, in Ann. NY Acad. Sc. 759, 
   ed. B\"oringer H, Morfill G.E. \\
   \mbox{  }\mbox{  }\& Tr\"umper J.E. (New York,  NY Acad. Sc.)\\
Nishikawa K, Koide S., Sakai J., Christodoulou D.M., Sol H.,
   Mutel R.L., 1997, ApJ, L45\\
Pearson T.J., Unwin S.C., Cohen M.H., Linfield R.P., 
    Readhead A.C.S., Seielstad G.A., \\
    \mbox{  }\mbox{  }Simon, R.S. \& Walker, R.C., 1981, Nature, 290, 365\\
Rees J.M. \& Meszaros P., 1994, ApJ, 430, L93\\
Thomson R.C., Mackay C.D. \& Wright A.E., 1993, Nature, 365, 133\\
Roe P.L., 1981, J. Comput. Phys., 43, 357\\
Ruggeri T. Strumia A., 1981, J. Math. Phys., 22, 1824\\
van Putten M.H.P.M., 1991, Commun. Math. Phys, 141, 63\\
\mbox{ }\mbox{ }--, 1993a, J Comput. Phys., 105(2), 339\\
\mbox{ }\mbox{ }--, 1993b, ApJ Lett., 408, L21\\
\mbox{ }\mbox{ }--, 1994a, Internat. J Bifur. \& Chaos, 4(1), 57\\
\mbox{ }\mbox{ }--,~1994b, in Proc. C. Lanczos~Int.~Centenary. Conf.,\\ 
	\mbox{  }\mbox{  }Moody~C., ed. Plemmons~R.,Brown~D.~and~
	Ellison D. \mbox{  }\mbox{  }(Philadelphia,  SIAM), p449\\
   \mbox{ }\mbox{ }--, 1994c, in Proc.~VIIth Conf. Waves and Stability in \\
	\mbox{  }\mbox{  }Continuous media, Bologna, 1993,
	ed. T. Ruggeri (World Scientific, Italy) \\
\mbox{ }\mbox{ }--, 1994d, Phys. Rev. D. 50(10):6640\\	
\mbox{ }\mbox{ }--, 1995, SIAM J Numer. Anal., 32(5), 1504\\
\mbox{ }\mbox{ }--, 1996, ApJ, 467, L21\\
van Putten M.H.P.M. \& Eardley D.M., 1996, Phys. Rev. D., 53, 3056\\
Wen, L., Panaitescu, A., \& Laguna, P., 1997, ApJ, 486, 919
}
\end{document}